\documentclass[12pt]{article}
\usepackage{amssymb}
\newtheorem{theorem}{Theorem}

\newif\iffigs\figstrue

\DeclareFontFamily{U}{rsf}{}
\DeclareFontShape{U}{rsf}{m}{n}{
  <5> <6> rsfs5 <7> <8> <9> rsfs7 <10-> rsfs10}{}
\DeclareMathAlphabet\Scr{U}{rsf}{m}{n}

\def\pplogo{\vbox{\kern-\headheight\kern -43pt
\halign{##&##\hfil\cr&{
\ppnumber}\cr\rule{0pt}{2.5ex}&\ppdate\cr}
}}
\makeatletter
\def\ps@firstpage{\ps@empty \def\@oddhead{\hss\pplogo}%
  \let\@evenhead\@oddhead 
a left-hand page
}
\def\maketitle{\par
 \begingroup
 \def\thefootnote{\fnsymbol{footnote}}
 \def\@makefnmark{\hbox{$^{\@thefnmark}$\hss}}
 \if@twocolumn
 \twocolumn[\@maketitle]
 \else \newpage
 \global\@topnum\z@ \@maketitle \fi\thispagestyle{firstpage}\@thanks
 \endgroup
 \setcounter{footnote}{0}
 \let\maketitle\relax
 \let\@maketitle\relax
 \gdef\@thanks{}\gdef\@author{}\gdef\@title{}\let\thanks\relax}
\makeatother

\begin{document}


\vspace{.1in}

\begin{center}
{\Large\bf REMARKS ON THE SPECTRUM AND TRUNCATED HEAT KERNEL OF THE BTZ BLACK HOLE}
\end{center}
\vspace{0.1in}
\begin{center}
{\large A. A. Bytsenko $^{(a)}$\footnote{ abyts@uel.br}, M. E. X.
Guimar\~aes $^{(b)}$ \footnote{marg@unb.br} and F. L. Williams
$^{(c)}$}
\footnote{williams@math.umass.edu}\\
\vspace{7mm}
$^{(a)}$ {\it Departamento de F\'{\i}sica, Universidade Estadual de
Londrina\\
Caixa Postal 6001, Londrina-Paran\'a, Brazil} \vspace{5mm}\\
$^{(b)}$ {\it Departamento de Matem\'atica, Universidade
de Bras\'{\i}lia, DF, Brazil}\vspace{5mm}\\
$^{(c)}$
{\it Department of Mathematics and Statistics,
University of Massachusetts \\
Lederle Graduate Research Tower 710 North Pleasant Street \\
Amherst, MA 01003, USA}\\
\end{center}
\vspace{0.1in}
\begin{center}
{\bf Abstract}
\end{center}
Using an orbifold description of the Euclidean BTZ black hole, we
show that there is a special relation between the  spectrum and
the  truncated heat kernel of this black hole with the
Patterson-Selberg zeta function.

\vspace{0.5cm}



\section{Introduction}

The discovery of black hole solutions in three-dimensional gravity
is a promising new area for the analysis of problems posed in the
four-dimensional case. We do not have yet a consistent and
complete theory of four-dimensional quantum gravity, but
nevertheless a large number of interesting issues have been
investigated using the three-dimensional analogs.  Some of them,
related to black hole thermodynamics for instance, deal with the
origin of entropy, the information loss paradox, and the validity
of the area law.
 Recently, three-dimensional gravity has been studied in more details.
 Despite of the simplicity of the three-dimensional
  case (with no propagating gravitons, for example), there is a common belief that it
  deserves attention as a useful laboratory for four-dimensional problems.

 In this paper we pursue this idea and we consider
 the Ba\~nados, Teitelboim, Zanelli (hereafter called BTZ) black hole \cite{Banados} because its geometric structure allows for exact computations since
  its Euclidean form is locally isomorphic to constant curvature hyperbolic three-space \cite{Bytsenko}.
We present a further discussion of the BTZ black hole in this
paper, as regards to its spectrum and truncated heat kernel.  We
find, in fact, that the latter two can be related via a
Patterson-Selberg zeta function \cite{Patterson}.  The contents of
the paper are as follows:  In Section 2 we describe the spectrum
of the cyclic, Kleinian group that defines the Euclidean black
hole as an orbifold.  In Section 3 we compute the truncated heat
trace and we show how to relate it to a Patterson zeta function.
Finally, in Section 4 we conclude with brief remarks.

\section{\protect \centering BTZ spectrum}
The Euclidean BTZ black hole has an orbifold description
$B_{\Gamma_{(a,b)}}=\Gamma_{(a,b)}\backslash H^{3}$ for suitable
parameters $a>0$, $b\geq 0$ (which we will specify later), where
$H^{3}=\{(x,y,z)\in\mathbb{R}^3\mid z>0\}$ is hyperbolic 3-space and
$\Gamma_{(a,b)}\subset SL(2,\mathbb{C})$ is a cyclic group of
isometries.  $B_{\Gamma_{(a,b)}}$ is a solution of the Einstein
equations
\begin{equation}
R_{ij}-\frac{1}{2}g_{ij}R_{g}-\Lambda g_{ij}=0
\end{equation}
with {\it negative} cosmological constant $\Lambda$, hyperbolic metric
\begin{equation}
ds^{2}=\frac{\sigma^{2}}{z^2}(dx^2+dy^2+dz^2)\label{eq: hypmetric}
\end{equation}
for $\sigma=(-\Lambda)^{-\frac{1}{2}}$, and constant scalar
curvature $R_{g}=\frac{6}{\sigma^2}=-6\Lambda$. The original BTZ
metric in coordinates $(r,\phi, \tau)$ can indeed be transformed
to that in (\ref{eq: hypmetric}) by a specific change of variables
$(r, \phi, \tau)\rightarrow (x,y,z)$; see for instance
\cite{Carlip, Bytsenko1, Perry}.  It is, in fact, the periodicity
in the Schwarzschild variable $\phi$ that allows for the above
orbifold description. In fact the parameters $a, b$ are given as
follows. For $M>0$, $J\geq 0$ the black hole mass and angular
momentum, and for $r_{+}>0, r_{-}\in i\mathbb{R}$ ($i^2=-1$) the
outer and inner horizons given by
\begin{eqnarray}
r_{+}^2&=&\frac{M\sigma^2}{2}\left[1+\left(1+\frac{J^2}{M^2\sigma^2}\right)^{\frac{1}{2}}\right]\,,\\
r_{-}&=&-\frac{\sigma Ji}{2r_{+}}\,,\label{eq: rs}
\end{eqnarray}
one obtains
\begin{equation}
a:=\pi r_{+}/\sigma,
b:=\pi\left|r_{-}\right|/\sigma. \label{eq:
ab}
\end{equation}
$\Gamma_{(a,b)}$ is defined to be the cyclic
subgroup of $G=SL(2,\mathbb{C})$ with generator
\begin{eqnarray}\label{eq: gammaGammaab}\gamma_{(a,b)}&:=&\left[
\begin{array}{cc}e^{a+ib} & 0 \\ 0 &
e^{-(a+ib)}\end{array}\right]:\\
\Gamma_{(a,b)} &:=&\{\gamma^n_{(a,b)}\mid n\in\mathbb{Z}\}\,.
\end{eqnarray} The Riemannian volume element $dv$
corresponding to (\ref{eq: hypmetric}) is given by
\begin{equation}dv=\frac{\sigma^3}{z^3}dxdydz.\label{eq: dv}\end{equation}
One knows that a fundamental domain $F_{(a,b)}$ for the action of $\Gamma_{(a.b)}$ on $H^3$ is given by
\begin{equation}F_{(a,b)}=\{(x,y,z)\in H^3 \mid 1<x^2+y^2+z^2<e^{2a}\}.\label{eq: Fab}\end{equation}
It follows that $\Gamma_{(a,b)}$ is a {\it Kleinian} subgroup of $G$ - i.e.
\begin{equation}
vol\left(F_{(a,b)}\right)=\int_{F_{(a,b)}}dv=\infty.
\end{equation}
Since $F_{(a,b)}$ has an infinite hyperbolic volume, the usual
spectral theory for the Laplacian $\Delta_{\Gamma_{(a,b)}}$ of
$B_{\Gamma_{(a,b)}}$ does {\it not} apply - as it does for finite
volume orbifolds.  We outline, briefly, a suitable spectral
analysis of $-\Delta_{\Gamma_{(a,b)}}$ where a key notion is that
of {\it scattering resonances}.  These replace the role of the
eigenvalues of the Laplacian in the infinite volume case, and are
the $s_{mnj}^{\pm}$ given in definition (\ref{eq: smnj}) below,
which we therefore refer to as the {\it BTZ spectrum}.

Henceforth we shall write $\Gamma$ for $\Gamma_{(a,b)}$.  Using (\ref{eq: hypmetric}), one notes that $\Delta_\Gamma$ is given by
\begin{equation}
\Delta_\Gamma=
\frac{1}{\sigma^2}\left[z^2\left(\frac{\partial^2}{\partial
x^2}+\frac{\partial^2}{\partial y^2} +\frac{\partial^2}{\partial
z^2}\right)-z\frac{\partial}{\partial z}\right].
\end{equation}
The
space of square-integrable functions on the black hole
$B_\Gamma=\Gamma\backslash H^3$, with respect to the measure $dv$ in
(\ref{eq: dv}), has a nice orthogonal decomposition
\begin{equation}
L^2\left(B_\Gamma,dv\right)=\sum_{m,n\in\mathbb{Z}}\oplus H_{mn}
\end{equation}
with Hilbert space isomorphisms $H_{mn}\simeq
L^2\left(\mathbb{R}^{+},dt\right)$ (for $\mathbb{R}^{+}$ the space
of positive real numbers), and a spectral decomposition
\begin{equation}
-\sigma^2\Delta_\Gamma\simeq\sum_{m,n\in\mathbb{Z}}\oplus L_{mn}
\end{equation}
where the
\begin{equation}
L_{mn}=-\frac{d^2}{dt^2}+1+V_{mn}(t)
\label{L}
\end{equation}
are the Schr\"{o}dinger operators with P\"{o}schel-Teller
potentials
\begin{equation}
V_{mn}(t)=\left(k_{mn}^2+\frac{1}{2}\right){\rm sech}^2t+\left(m^2-\frac{1}{4}\right){\rm cosh}^2t\end{equation}
for
\begin{equation}
k_{mn}:= -\frac{mb}{a}+\frac{\pi n}{a}.\label{eq: kmn}
\end{equation}
For details of this and the following remarks the reader can
consult \cite{Perry, Guiloppe, Sjostrand}, for example.
In particular, the proof of equation (\ref{L}) is given in Section 3 of \cite{Perry} - in equations (3.10)-(3.21) there.
The Schr\"{o}dinger equation
\begin{equation}
\Psi ''(x)+[E-V_{mn}(x)]\Psi(x)=0,
\end{equation}
which is the same as the eigenvalue problem $L_{mn}\Psi=k^2\Psi$ for
$E=k^2-1$, has a known solution $\Psi^+(x)$ (in terms of the
hypergeometric function) with asymptotics
\begin{equation}
\Psi^+(x)\sim
\frac{e^{ikx}}{T_{mn}(k)}+
\frac{R_{mn}(k)}{T_{mn}(k)}e^{-ikx},
\end{equation}
for
{\it reflection} and {\it transmission} coefficients
$T_{mn}(k)$, $R_{mn}(k)$.  For $s$ defined by $k=i(1-s)$ one can
form the {\it scattering matrix}
\begin{equation}
\left[\mathcal{R}_{mn}(s)\right]:=\left[R_{mn}(k)\right]
\end{equation}
of $-\Delta_\Gamma$, whose entries are quotients of gamma
functions with ``trivial poles" $s=1+j$,
$j=0,1,2,3,\mathellipsis,$ and non-trivial poles
\begin{equation}
\label{eq: smnj}s_{mnj}^{\pm}:= -2j-\left|m\right|\pm i\left|k_{mn}\right|
\end{equation}
for $k_{mn}$ in (\ref{eq: kmn}), $j=0,1,2,3,\mathellipsis$; also see
(\ref{eq: ab}).  The $s^{\pm}_{mnj}$ are the scattering resonances
that we referred to earlier.

For later purposes it is convenient to set
\begin{equation}
\label{eq: ltheta}l := 2a=2\pi r_{+}/\sigma, \theta := 2b=2\pi\left|r_{-}\right|/\sigma;
\end{equation}
see (\ref{eq: ab}).  For $n,k_{1},k_{2}\in\mathbb{Z},
k_{1},k_{2}\geq 0$ define a corresponding complex number
$\zeta_{n,k_{1},k_{2}}$ by
\begin{equation}
\label{eq: zetank1k2}\zeta_{n,k_{1},k_{2}}:= -\left(k_{1}+k_{2}\right)+i\left(k_{1}-k_{2}\right)\frac{\theta}{l}+\frac{2\pi  in}{l}.
\end{equation}
The $\zeta_{n,k_{1},k_{2}}$ turns out to be the zeros of a zeta
function $Z_\Gamma(s)$
 that we will introduce in the next section, where we also relate $Z_\Gamma(s)$ to the heat kernel of $B_\Gamma$.
It is a simple, but remarkable, fact that the set of scattering
poles in (\ref{eq: smnj}) {\it coincides} with the zeta zeros in
(\ref{eq: zetank1k2}), as is easily verified.  Thus encoded in
$Z_\Gamma(s)$ is the spectrum of a BTZ black hole.

\section{\protect \centering  BTZ heat kernel and zeta function}

In \cite{Carlip}, the heat kernel trace (integration over the
fundamental domain $F=F_{(a,b)}$ in (\ref{eq: Fab}) along the
diagonal) was calculated for the non-spinning black hole - the
case $b=0$ in (\ref{eq: ab}).  In this section we indicate how to
perform  that calculation for the general a spinning black hole
for arbitrary values of $b$.  An alternate computation appears in
\cite{Mann}. We, moreover, relate the result (which is not done in
\cite{Bytsenko1}, nor in \cite{Mann}) to a zeta function
$Z_\Gamma(s)$ whose zeros comprise the BTZ spectrum, as mentioned
in Section 1.

The calculation is carried out conveniently with spherical
coordinates: for $\rho\geq 0, 0\leq\theta\leq 2\pi, 0\leq\phi\leq
\frac{\pi}{2}, x=\rho \sin{\phi} \cos{\theta}, y=\rho \sin{\phi}
\sin{\theta}, z=\rho \cos{\phi}$. We have also used $\theta$ to
denote $2b$ in definition (\ref{eq: ltheta}).  This dual use of
notation should be no cause for confusion.  For $p=(x,y,z)\in H^3$,
its image (=its $\Gamma$-orbit) under the quotient map
$H^3\rightarrow\Gamma\backslash H^3=B_\Gamma$ will be denoted by
$\tilde{p}$.  $d(p_{1},p_{2})$ will denote the hyperbolic
distance between two points
$p_{1},p_{2}=(x_{1},y_{1},z_{1}),(x_{2},y_{2},z_{2})$ in $H^3$:
\begin{equation}\label{eq: coshdp1p2}\cosh{ d(p_{1},p_{2})} := 1+\frac{(x_{1}-x_{2})^2+(y_{1}-y_{2})^2+(z_{1}-z_{2})^2}{2z_{1}z_{2}}.
\end{equation}
The heat kernel $K^\Gamma_t$ ($t>0$) of $B_\Gamma$ is obtained by averaging over $\Gamma$ the heat kernel $K_t$ of $H^3$:
\begin{eqnarray}\label{eq: KGammat}K^\Gamma_t\left(\widetilde{p_{1}},\widetilde{p_{2}}\right)&=&\sum_{n\in\mathbb{Z}}
K_t\left(p_{1},\gamma^n p_{2}\right) \\
&=&\sum_{n\in\mathbb{Z}}\frac{e^{-t-d\left(p_{1},\gamma^n
p_{2}\right)^2/ 4t}d\left(p_{1},\gamma^n p_{2}\right)}{(4\pi
t)^\frac{3}{2}\sinh{d\left(p_{1},\gamma^n
p_{2}\right)}},
\end{eqnarray}
given definition (\ref{eq:
gammaGammaab}), where we write $\gamma$ for $\gamma_{(a,b)}$.  As
above we will write $F$ for $F_{(a,b)}$ in (\ref{eq: Fab}). Finally,
let $K_t^{*\Gamma}(\widetilde{p_{1}},\widetilde{p_{2}})$ denote the
{\it truncated} heat kernel of $B_\Gamma$, defined by
restricting the sum over $\mathbb{Z}$ in (\ref{eq: KGammat}) to the
{\it non-zero} integers $n$.  We can now prove the following
theorem for the trace of
$K_t^{*\Gamma}(\widetilde{p_{1}},\widetilde{p_{2}})$.
\begin{theorem} For the volume element $dv$ in (\ref{eq: dv}), and the theta-function
\begin{eqnarray}\label{eq: ThetaGamma}\Theta_{\Gamma}(t)& := &
\frac{l}{8\sqrt{4\pi t}}
\sum_{n\in\mathbb{Z}-\{0\}}\frac{e^{-\left(t+\frac{n^2l^2}{4t}\right)}}{[\sinh^2\left(\frac{ln}{2}\right)+
\sin^2\left(\frac{\theta n}{2}\right)]}
\\
&=&\frac{l}{4\sqrt{4\pi t}}\sum_{n=1}^\infty
\frac{e^{-\left(t+\frac{n^2l^2}{4t}\right)}}
{[\sinh^2\left(\frac{ln}{2}\right)+\sin^2\left(\frac{\theta
n}{2}\right)]},
\end{eqnarray}
for $t>0$, see (\ref{eq:
ltheta}), one has that
\begin{equation}
\int\int\int_{F} K_{t}^{*\Gamma}(\widetilde{p},\widetilde{p})dv=2\sigma^3\Theta_{\Gamma}(t).
\end{equation}
\end{theorem}
{\it Proof.}
For $n\in\mathbb{Z}-\{0\}$,
let $r_n := 2\pi n\left|r_{-}\right| /\sigma$. In terms of the
above spherical coordinates, the action of $\Gamma$ on $H^3$ (which
appears in particular in definition (\ref{eq: KGammat})) is given by
$\gamma^n(x,y,z)=(x',y',z')$ for
$x'=e^{nl}(\rho\sin{\phi})\cos(\theta+r_n)$,
$y'=e^{nl}(\rho\sin{\phi})\sin(\theta+r_n)$,
$z'=e^{nl}\rho\cos{\phi}$, with $l=2\pi r_{+}/\sigma$ in (\ref{eq:
ltheta}).  Then one can compute that
$(x-x')^2+(y-y')^2+(z-z')^2=\rho^2(\sin^2{\phi})[1-2e^{nl}\cos{r_n}+(e^{nl})^2]+\rho^2(\cos^2{\phi})[1-e^{nl}]^2=
\rho^2[1-e^{nl}]^2-2\rho^2e^{nl}(\sin^2{\phi})(\cos{r_n}-1)$. For
$N := e^l$, $b_n :=[1-N^n]^2/2N^n$, $d_n :=
d(p,\gamma^n p)$, and $\widetilde{b_n} :=(1+b_n-\cos{r_n})$,
one obtains from (\ref{eq: coshdp1p2}) that
\begin{eqnarray}
\label{eq: circle} \cosh{d_n} & = &
1+\frac{[1-N^n]^2-2N^n(\sin^2{\phi})(\cos{r_n}-1)}{2(\cos^2\phi)N^n}
\nonumber \\
& =
 & 1+b_n\sec^2\phi-(\tan^2\phi)(\cos{r_n}-1)=\cos{r_n}+(b_n+1-\cos{r_n})\sec^2\phi
\nonumber \\
& = & \cos{r_n}+\widetilde{b_n}\sec^2\phi \, ,
\end{eqnarray}
which is independent of the other spherical coordinates $\rho$ and
$\theta$.
 Note that, by definition
 \begin{equation}\label{eq: 1plusbnbntilde}
1+b_n=\cosh{nl}\,,\,\,\,\, \widetilde{b_n}=\cosh{nl}-\cos{r_n}\,.
\end{equation}
As $dv=\sigma^3\left(\sin\phi\right)/\rho\cos^3\phi d\rho d\theta d\phi$\,, commutation of integration and summation
(where again the summation in (\ref{eq: KGammat}) is restricted to $\mathbb{Z}-\{0\}$ for $K_t^{*\Gamma}\left(\widetilde{p},\widetilde{p}\right)$ ) gives
 \begin{equation}\label{eq: intintintKtGamma2}\int\int\int_F K_t^{*\Gamma}\left(\widetilde{p},\widetilde{p}\right)dv=
 \frac{e^{-t}\sigma^3}{\left(4\pi t\right)^\frac{3}{2}}
 \sum_{n\neq 0}I_n\end{equation}
 for
 \begin{equation}I_n=\int_0^{\pi/2}\int_0^{2\pi}\int_1^{N}
 \frac{e^{-d_n^2/4t}d_n\left(\sin\phi\right)}
 {\left(\sinh d_n\right)\rho\cos^3{\phi}}d\rho d\theta d\phi,\end{equation}
 where by (\ref{eq: circle}), $d_n=d_n\left(\phi\right)$ depends only on $\phi$, and not on $\theta, \rho$.  Therefore
 \begin{equation}I_n=2\pi\left(\log{N}\right)\int_0^{\pi/2}\frac{e^{-d_n^2\left(\phi\right)/4t}d_n\left(\phi\right)\left(\sin\phi\right)}
 {\left(\sinh d_n\left(\phi\right)\right)\cos^3{\phi}}d\phi.\end{equation}
 Differentiating equation (\ref{eq: circle}) with respect to $\phi$ and using the change of variables $u=d_n\left(\phi\right)$
\\
we get
that$\left(\sinh{d_n\left(\phi\right)}\right)d_n'\left(\phi\right)=\widetilde{b_n}2\sec^2\phi\tan\phi=
2\widetilde{b_n}\sin\phi/\cos^3\phi \Rightarrow
du=d_n'\left(\phi\right)d\phi=
\\
2\widetilde{b_n}\left(\sin\phi\right)/\left(\sinh{d_n\left(\phi\right)}\right)\cos^3\phi d\phi\Rightarrow$
\begin{eqnarray}
I_n&=&2\pi\left(\log{N}\right)\int_{d_n(0)}^{d_n(\pi/2)}e^{-u^2/4t}\frac{u}{2\widetilde{b_n}}du\\
&=&-2\pi\left(\log{N}\right)\frac{t}{\widetilde{b_n}}[e^{-u^2/4t}]_{d_n(0)}^{d_n\left(\pi/2\right)}\,.
\end{eqnarray}
By (\ref{eq: circle}) and (\ref{eq: 1plusbnbntilde}),
$d_n\left(\phi\right)=\cosh^{-1}\left(\cos{r_n}+\widetilde{b_n}\sec^2\phi\right)=\log
\left[\cos{r_n}+\widetilde{b_n}\sec^2\phi
+\right.\\
\left.\sqrt{\left(
\cos{r_n}+\widetilde{b_n}\sec^2\phi\right)^2-1}\right]\Rightarrow
d_n(0)=\log[\cosh{nl}+\left|\sinh{nl}\right|]=\left|n\right|l$. Also
$d_n\left(\frac{\pi}{2}\right)=\infty$, and we see that $I_n=({2\pi
lt/\widetilde{b_n}})e^{-n^2l^2/4t}$ (as $\log{N} :=
l$)$\Rightarrow$ (by (\ref{eq: 1plusbnbntilde}), (\ref{eq:
intintintKtGamma2}))
\begin{eqnarray}\int\int\int_F
K_t^{*\Gamma}\left(\widetilde{p},\widetilde{p}\right)dv&=&\frac{\sigma^3
2\pi l t}{\sqrt{4\pi t}(4\pi t)}\sum_{n\neq
0}\frac{e^{-t-n^2l^2/4t}}{\widetilde{b_n}} \\
&=&\frac{\sigma^3 l}{\sqrt{4\pi
t}}\sum_{n=1}^\infty\frac{e^{-t-n^2l^2/4t}}{\left[\cosh(nl)+\cos(r_n)\right]}\\
&=&\frac{\sigma^3 l}{2\sqrt{4\pi
t}}\sum_{n=1}^\infty\frac{e^{-t-n^2l^2/4t}}{\left[\sinh^2
\left(\frac{ln}{2}\right)+\sin^2\left(\frac{r_n}{2}\right)\right]}\,,
\end{eqnarray}
which by the definition $r_n=2\pi
n\left|r_{-}\right|/\sigma=n\theta$ (see (\ref{eq: ltheta}))
concludes the proof.\, $\square$

The following zeta function has been attached to the BTZ black hole $B_\Gamma:$
\begin{equation}
Z_\Gamma(s) :=\prod_{\stackrel{k_1,k_2\geq
0}{k_1,k_2\in\mathbb{Z}}}^\infty[1-(e^{i\theta})^{k_1}(e^{-i\theta})^{k_2}e^{-(k_1+k_2+s)l}]\,,
\label{zeta}
\end{equation}
again for $l,\theta$ in (\ref{eq: ltheta}); see \cite{Perry}, \cite{Williams}.
\footnote{
Strictly speaking the zeta function in definition (\ref{zeta})
does have a connection to BTZ black hole entropy. Namely, as discussed in \cite{Perry}, \cite{Williams}, quantum corrections to the classical Beckenstein-Hawking entropy can be expressed as a special value of the logarithm of this function.}
$Z_\Gamma(s)$
is an entire function of $s$, whose zeros are precisely the complex
numbers $\zeta_{n,k_1,k_2}$ given in (\ref{eq: zetank1k2}), and
whose logarithmic derivative is given by
\begin{equation}\label{eq: ZGammaprimeoverZGamma}\frac{Z_\Gamma '(s)}{Z_\Gamma(s)}=
\frac{l}{4}\sum_{n=1}^{\infty}\frac{e^{-nl(s-1)}}{[\sinh^2\left(\frac{ln}{2}\right)
+\sin^2\left(\frac{\theta n}{2}\right)]}
\end{equation} for $Res >
0$.  In Section 1, we connected $Z_\Gamma(s)$ with the BTZ spectrum.
$Z_\Gamma(s)$ is also connected with the theta function
$\Theta_\Gamma(t)$ in (\ref{eq: ThetaGamma}), and hence with the
heat kernel $K_t^{*\Gamma}$ (by Theorem 1) via the following
theorem, which follows easily from (\ref{eq: ZGammaprimeoverZGamma})
by commuting integration and summation in (\ref{eq:
ThetaGamma}):
\begin{theorem}
For $Re\,s>1$
\begin{equation}
\int_0^\infty e^{-s(s-2)t}\Theta_\Gamma(t)dt=\frac{1}{2(s-1)}
\frac{Z'_\Gamma(s)}{Z_\Gamma(s)}\,\,.
\end{equation}
\end{theorem}

{\it Proof.} The proof of Theorem 2 relies on the Laplace
transform formula
\begin{equation}
\int_0^{\infty}t^{-\frac{1}{2}}e^{-\frac{a}{4t}}e^{-pt} dt =
\pi^{\frac{1}{2}} p^{-\frac{1}{2}}e^{-(ap)^{\frac{1}{2}}}
\end{equation}
for $Re\,a\geq 0$, $Re\,p>0$.  Using definition (27), one gets
\begin{equation}
\int_0^{\infty}e^{-s(s-2)t}\Theta_\Gamma(t)dt =
\frac{l}{4\sqrt{4\pi}} \sum_{n=1}^\infty
\frac{\int_0^\infty
e^{-s(s-2)t -\left(t+\frac{n^2l^2}{4t}\right)}
\frac{dt}{\sqrt{t}}}
{\sinh^2\left(\frac{ln}{2}\right)+\sin^2\left(\frac{\theta
n}{2}\right)},
\end{equation}
where $e^{-s(s-2)t -\left(t+\frac{n^2l^2}{4t}\right)} \equiv
e^{-t(s-1)^2 -\frac{n^2l^2}{4t}}$.  For the choices $a=n^2l^2$,
$p=(s-1)^2$ in (42), the second integral in (43) therefore assumes
the value $\left(\sqrt{\pi}/(s-1)\right)e^{-nl(s-1)}$, first for
$s>1$ and then for $Re\,s>1$ by analytic continuation. The first
integral in (42) then is the sum
\begin{equation}
\frac{l}{4\cdot 2(s-1)}\sum_{n=1}^\infty
\frac{e^{-nl(s-1)}}{\sinh^2\left(\frac{ln}{2}\right)+\sin^2\left(\frac{\theta
    n}{2}\right)} = \frac{1}{2(s-1)}\frac{Z_\Gamma '(s)}{Z_\Gamma (s)},
\end{equation}
by equation (40), for $Re\,s>1$, which establishes Theorem 2.
$\square$

\section{\protect \centering  Concluding remarks}

The main goal of our paper was to show that, using an orbifold
description of the Euclidean BTZ black hole, one can obtain an
interesting relation between the BTZ spectrum and its truncated
heat kernel and the Patterson-Selberg zeta function. This relation
is provided by Theorem 2 in our paper. Zeta-function methods are a
powerful tool to obtain the spectral information of operators.
With our result, we hope to give a first step towards
understanding deeply the thermodynamics and statistical properties
of the BTZ black holes and other (maybe, more complex)
three-dimensional systems.

\section*{\protect \centering  Acknowledgments}

A. A. Bytsenko and M. E. X. Guimar\~aes would like to thank the
Conselho Nacional de Desenvolvimento Cient\'{\i}fico e
Tecnol\'ogico (CNPq-MCT) for partial financial support.



\begin{thebibliography}{77}


\bibitem{Banados}
M. Ba\~nados, C. Teitelboim and J. Zanelli, The black Hole in Three Dimensional Space Time,
{\it Phys. Rev. Lett.} {\bf 69} (1992) 1849-1851 [hep-th/9204099].

\bibitem{Bytsenko}
A. A. Bytsenko, M. E. X. Guimar\~{a}es and R. Kerner, Orbifold
compactification and solutions of M-theory from Milne spaces, {\it
Eur. Phys. J. C } {\bf 39} (2005) 519-524 [hep-th/0501008].

\bibitem{Patterson}
S. Patterson, The Selberg zeta function of a Kleinian group - from Number Theory, Trace Formulas, and Discrete Groups:  Symposium in Honor of Alte Selberg, Academic Press (1989).

\bibitem{Carlip}
S. Carlip and C. Teitelboim, Aspects of black hole quantum mechanics and thermodynamics in 2+1 dimensions, {\it Phys. Rev. D } {\bf 51} (1995) 622-631 [gr-qc/9405070].

\bibitem{Bytsenko1}
A. A. Bytsenko, L. Vanzo and S. Zerbini, Quantum correction to the
entropy of the (2+1)-dimensional black hole, {\it Phys. Rev. D}
{\bf 57} (1998) 4917-4924 [gr-qc/9710106].

\bibitem{Perry}
P. Perry and F. Williams, Selberg zeta function and trace formula for the BTZ black hole, {\it Internat. J. of Pure and Applied Math.} {\bf vol 9} (2003) 1-21.

\bibitem{Guiloppe}
L. Guillop\'{e} and M. Zworski, Upper bounds on the number of resonances for non-compact Riemann surfaces, {\it J. of Funct. Analysis} {\bf 129} (1995) 364-389.

\bibitem{Sjostrand}
J. Sj\"{o}strand and M. Zworski, Lower bounds on the number of scattering poles, {\it Comm. Partial Diff. Equations} {\bf 18} (1993) 847-857.

\bibitem{Mann}
R. Mann and S. Solodukhin, Quantum scalar field on a three-dimensional (BTZ) black hole instanton: heat kernels, effective action, and thermodynamics, {\it Phys. Rev. D} {\bf 55} (1997) 3622-3632.

\bibitem{Williams}
F. Williams, A zeta function for the BTZ black hole, {\it Internat. J. Modern Physics A} {\bf 18}  (2003) 2205-2209.




\end{thebibliography}
\end{document}